\documentclass[%
 aip,
 amsmath,amssymb,
 reprint,%
]{revtex4-1}

\usepackage{graphicx}
\usepackage{dcolumn}
\usepackage{bm}

\usepackage[utf8]{inputenc}
\usepackage[T1]{fontenc}
\usepackage{mathptmx}
\usepackage{setspace}

\def\AOM {acousto-optic modulator}
\def\BCA {boxcar averager}
\def\ECDL {extended cavity diode laser}
\def\FVC {frequency-to-voltage converter}

\def\MOT {magneto-optical trap}
\def\PMT {photo-multiplier tube}
\def\ppKTP {periodically poled potassium titanyl phosphate}  
\def\SNR {signal-to-noise ratio}

\def\VCO {voltage controlled oscillator}

\def\ARC {Australian Research Council} 

\newcommand{\degC}{$^{\circ}$C}

\newcommand{\uK}{$\mu$K}  
\newcommand{\us}{$\mu$s}

\newcommand{\fastT}{$^{1}S_{0}-\,^{1}P_{1}$}    
\newcommand{\coolingT}{$^{1}S_{0}-\,^{3}P_{1}$}  

\newcommand{\clockT}{$^{1}S_{0}-\,^{3}P_{0}$}

\newcommand{\si}{$\sim$}

\newcommand{\Yb}{$^{171}$Yb}	

\newcommand{\Ybthree}{$^{173}$Yb}

\begin{document}

\preprint{AIP/123-QED}

\title{Fourier transform detection of weak optical transitions with cyclic routines}

\author{ Jesse~S.~Schelfhout, Lilani~D.~Toms-Hardman and  John~J.~McFerran}
 \email{john.mcferran@uwa.edu.au}
\affiliation{ 
Department of Physics, University of Western Australia, 35 Stirling Highway, 6009 Crawley, Australia
}%


\date{\today}

\begin{abstract}
We demonstrate a means of detecting weak optical transitions in  cold atoms that undergo cyclic routines with high sensitivity.  The gain in sensitivity is made by  probing atoms on alternate cycles leading to a regular modulation of the ground state atom population when at the  resonance frequency.   The atomic transition is identified by conducting  a fast Fourier transform via algorithm or instrument. 
We find an enhancement of detection sensitivity compared to more conventional scanning methods of $\sim 20$ for the same sampling time, and can detect  clock lines with  fewer than $10^3$ atoms in a magneto-optical trap. 
  We apply the method to the  $(6s^{2})\,^{1}S_{0} -(6s6p)\,^{3}P_{0}$ clock transition in  $^{171}$Yb and $^{173}$Yb. The ac-Stark shift of this line in  $^{171}$Yb  is measured to be 0.19(3)\,kHz$\cdot$W$^{-1}\cdot$m$^2$ at 556\,nm. 
\end{abstract}

\maketitle

We investigate a technique  of detecting weakly allowed optical transitions that have poor \SNR\ (SNR) when using conventional  frequency sweep methods.  This is of  importance given the resurgence of isotopic shift spectroscopy used to search for 
beyond Standard-Model signatures~\cite{Fla2018,Cou2020,Sol2020}. Forbidden transitions in neutral atoms, such as the \clockT\  transition in group-II-like  atoms~\cite{Tak2005,Lud2015} 
may be used in such investigations. 
Knowledge of isotopic shifts is also relevant to many radiopharmaceutical  applications~\cite{Gre1990,Par2008,Kos2020}.
We demonstrate a means to increase the sensitivity of detecting weak transitions and apply it to the  clock transition in the composite fermions of ytterbium~\cite{Hoy2005, Hin2013}. 
The method is applicable to any  transition in  atoms that  undergo cyclic preparation before probing.
With regard to searching for atomic transitions, the technique is appropriate where the ratio of search  space in frequency, $\Delta f_S$, to transition linewidth, $\Delta \nu$,  is not excessively large.  For example,  where the transition frequency can be estimated using the knowledge of isotope shifts~\cite{Zin2002}, or in the case of photoassociation spectroscopy, estimated using a Le Roy$-$Bernstein type equation~\cite{Wei1999,Toj2006}.  
 The gain in detection sensitivity is made by  cycling the probe light and looking for a corresponding signal in the Fourier domain.  
  For example,  if we  refer to each atom loading and cooling cycle as one cycle, then for subsequent cycles the probe light  alternates between  on and off.  When on resonance, some fraction of  atoms will be excited to the upper state every second cycle.  This appears as a modulation of the fluorescence received by the photodetector, which is proportional to the number of atoms in the ground state. The strength of the modulation  is extracted through the FFT of a time record of the photoreceiver's signal.   We demonstrate the sensitivity of the method by  performing spectroscopy on the \clockT\ line in \Yb\ and \Ybthree, which have natural linewidths of approximately 40\,mHz~\cite{Por2004}.  
  A Fourier signal with $\sim10$\,dB SNR is observed for the \Yb\ clock line with $\sim10^3$ atoms. 
 We note that the technique should be well suited to searching for clock transitions in bosonic isotopes when held in an optical lattice trap~\cite{Pol2008,Aka2010,Bai2007}.  It could also be applied in the search for photoassociation resonances in ultra cold molecules~\cite{Yas2006,Zel2006,Jon2006,Nem2009}, for example, at predicted frequencies, where otherwise they are not apparent~\cite{Gut2018}.   

\begin{figure}[b!]
    \begin{center}
        \includegraphics[width=0.46\textwidth]{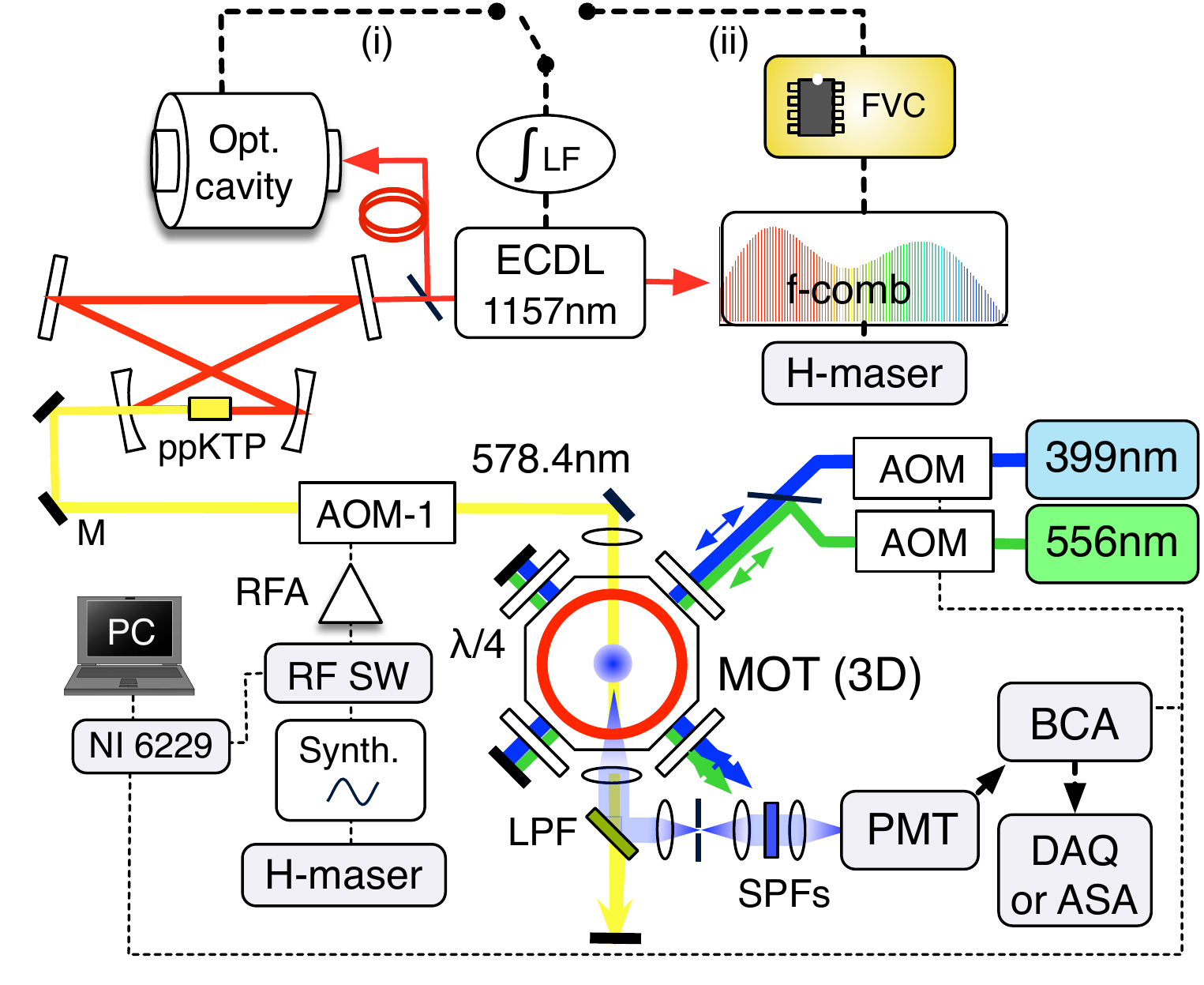}
        \caption{\small  An outline of the experimental layout for the \clockT\ line spectroscopy in Yb.   (i) and (ii) represent two means of stabilizing the 1157\,nm laser. AOM, \AOM; ASA, audio (FFT) spectrum analyser; BCA, \BCA;  ECDL, \ECDL; f-comb, frequency comb (in the near-IR); FVC,  \FVC; H, hydrogen; LPF, long-wavelength pass filter; MOT, \MOT; NI, National Instruments unit; PMT, \PMT; ppKTP, \ppKTP;  RFA, RF amplifier; SPF, short-wavelength pass filter; SW, switch.} %
             \label{ExpLayout}
    \end{center}
\end{figure}

 The experiment is comprised of  a \MOT\ (MOT), a frequency comb (MenloSystems FC1500), a hydrogen maser (KVARZ-75A), cooling lasers at 399\,nm and 556\,nm, and the  clock-line laser at 578\,nm to probe the \clockT\ transition.  The main components are shown in Fig.~\ref{ExpLayout}.   The source of ytterbium atoms is provided by an oven  at \si\,400\degC\ from which atoms effuse through an array of narrow collimation tubes orientated horizontally.  Some fraction of atoms pass through a narrow conduit downstream that separates the vacuum chamber into two main sections:  (i) the oven and (ii) the Zeeman slower and main chamber.

 The MOT operates in two stages; first with 399\,nm light acting on the \fastT\ transition, then 556\,nm light acting on the \coolingT\ transition to reduce the atomic temperature to \si50\,\uK~\cite{Comment1}. 
  A dual tapered Zeeman slower (crossing zero magnetic field) assists with  loading atoms into the MOT.   The 399\,nm light is generated by frequency doubling 798\,nm light from a Ti:sapphire laser in a resonant cavity.  The 556\,nm light is produced by use of a second frequency doubling cavity, where the incident 1112\,nm light originates from a fibre laser (NKT Photonics) that injection locks a semiconductor laser (EYP-RWL-1120).  The 399\,nm light is stabilized by  locking to the \fastT\ line in atoms progressing  to the Zeeman slower.  The 556\,nm light is stabilized by locking the sub-harmonic at 1112\,nm  to a mode of the frequency comb via a \FVC\ (FVC)~\cite{Rey2019}.   The cooling laser signals are controlled with \AOM s (AOMs) driven with amplified RF signals from \VCO s (VCOs).   
 
The clock line laser at 578\,nm is produced by use of a third frequency doubling cavity, where the master laser is an extended cavity diode laser (LD1001 Time-Base) and   H\"{a}nsch-Couillaud frequency stabilization
is used with the doubling cavity~\cite{Nen2016}.  The yellow light frequency is tuned with an additional AOM  (AOM-1 in Fig.~\ref{ExpLayout}) whose RF is set with a  synthesizer (Agilent E4428c) that is referenced to the H-maser (with accuracy in the $10^{-12}$ range).

The optical frequency of the 578\,nm light is evaluated with 
\begin{equation}
\label{EqComb}
\nu = 2 (n f_R - f_O \pm f_B) + f_\mathrm{aom}
\end{equation}
where $f_R$ and  $f_{O}$ are the frequency comb's mode spacing and offset frequency, respectively ($f_{O}=20$ MHz), $f_{B}$ is the  beat frequency,  $n$ is the mode number of the comb, and  $f_\mathrm{aom}$ is the RF drive frequency of the AOM   in the path of the 578\,nm light.  The sign of $f_{B}$ may vary depending on the isotope. 
The comb's repetition rate is controlled by mixing its fourth harmonic with a \si1.00\,GHz
signal that is the  sum of \si20\,MHz and 980.0\,MHz signals provided by a direct digital synthesizer  (DDS) and a dielectric resonator oscillator (DRO), respectively. Hence $f_R = (f_\mathrm{DDS} +f_\mathrm{DRO})/4$.   Both the DRO and DDS are  locked to a 10\,MHz signal from the H-maser. %
The DDS frequency is imposed by that of the 556\,nm cooling light.  


For the \Yb\ clock transition the 1157\,nm laser can be stabilized in one of two ways (summarized in Fig.~\ref{ExpLayout}): (i) with use of an ultrastable cavity and Pound-Drever-Hall lock, or (ii) by locking to the frequency comb  using the FVC technique.        The ultrastable cavity (Stable Laser Systems) and laser frequency stabilization have been described previously~\cite{Nen2016}.  The coefficient of thermal expansion is zero at $29.7(2)$\,\degC, and the second order thermal expansion coefficient is $6.6\times10^{-10}$\,K$^{-2}$  (for $\nu=259.1$\,THz). The laser when locked to the cavity drifts at a rate of 20.3\,mHz$\cdot$s$^{-1}$.  The free spectral range of the cavity is 1496.5210(1)\,MHz.  This is found by measuring the frequency separation between consecutive modes of the cavity  with the frequency comb and the 1157\,nm beat signal.  A more accurate determination is then found using the \Yb\ clock transition frequency. By locking to the 173167$^{th}$ mode of the cavity and offsetting the 578\,nm light  by   $-284.5$\,MHz with  AOM-1 the \clockT\ transition frequency can be reached.   The   \clockT\ transition for \Ybthree\ lies approximately midway between cavity modes, which  makes frequency offsetting with AOMs difficult.  
 To overcome this obstacle one could use offset sideband locking with the ultrastable cavity~\cite{Tho2008b}.  Instead, our approach  is to  lock the 1157\,nm laser to the frequency comb, which also gives flexibility in reaching the \clockT\ transition (rather than using a sequence of AOMs). %
There are some constraints because the comb is also used to stabilize the 556\,nm light frequency, and beat signals can only lie at frequencies set by the bandpass filters; here, 21\,MHz and 30\,MHz for the 1112\,nm and 1157\,nm beats, respectively.  %

\begin{figure}[h!]
    \begin{center}
        \includegraphics[width=0.38\textwidth]{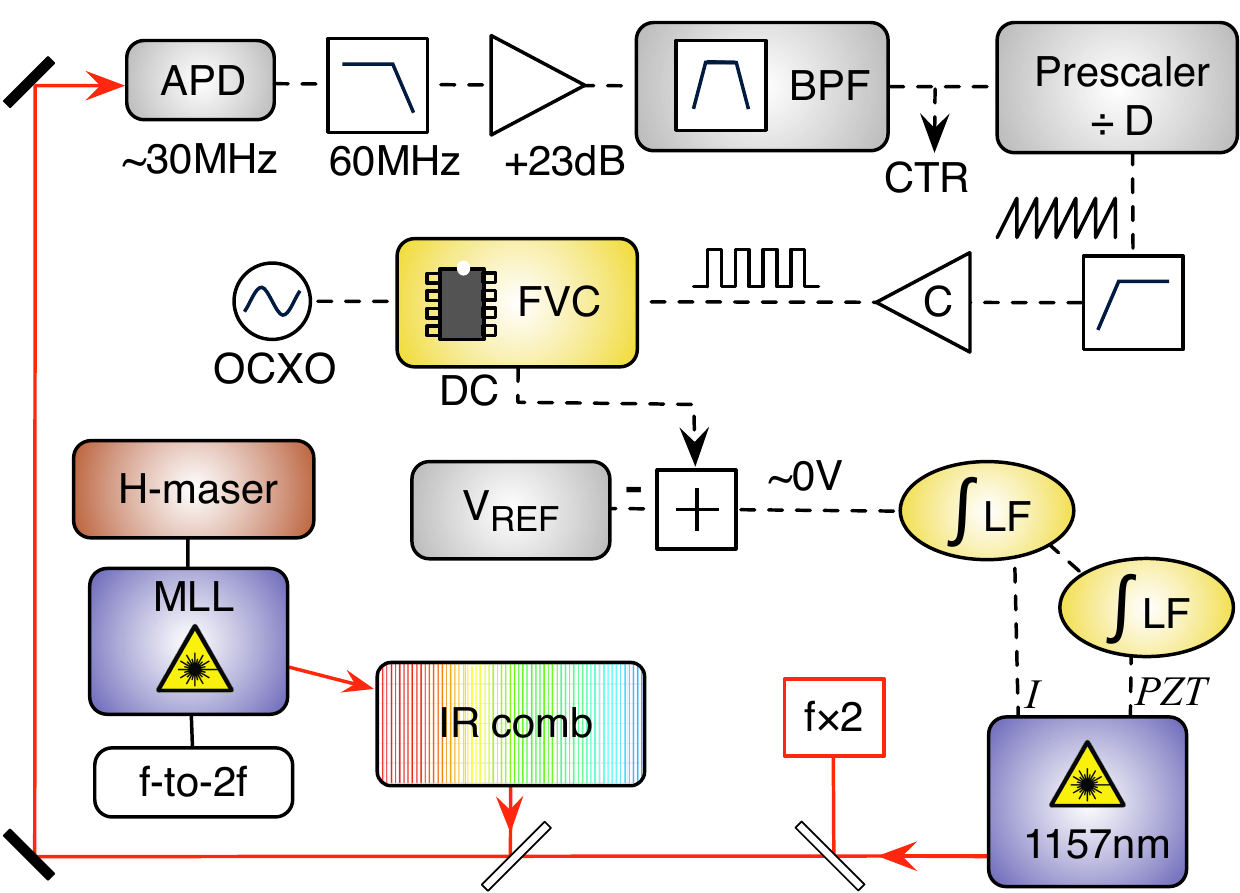}
        \caption{Laser frequency stabilization of the probe laser by use of a \FVC.   In the prescaler the frequency division is $D=128$. APD, avalanche photodiode;  BPF, band pass filter; C, comparator;  CTR, frequency counter;  LF, loop filter (integrator); MLL, mode-locked laser; OCXO, oven controlled crystal oscillator. PZT and $I$ represent electronic feedback to a piezo transducer and laser current, respectively. } 
             \label{FVClock}
    \end{center}
\end{figure}

The setup for the probe laser frequency stabilization by use of a FVC (Analog Devices AD652) is shown in Fig.~\ref{FVClock}.   After filtering and amplifying the beat signal from the avalanche photodiode it is divided in frequency  by 128 with a prescaler (Fujitsu MB506).  A comparator (LM360) regularizes the waveform  making it  suitable for the CMOS compatible FVC.  The AD652 is a synchronous VFC, implying that its transfer function is governed by an external clock, in this case an OCXO at 10\,MHz, divided by 4.   A laser servo based on this scheme has been shown to produce a fractional frequency instability of \si$7\times10^{-14}$ for $0.2<\tau<10^3$\,s, where $\tau$ is the integration time~\cite{Rey2019}. 
Apart from the flexibility provided by locking to the comb, the FVC-comb lock is also extremely robust.  

\begin{figure}[h!]
    \begin{center}
        \includegraphics[width=0.5\textwidth]{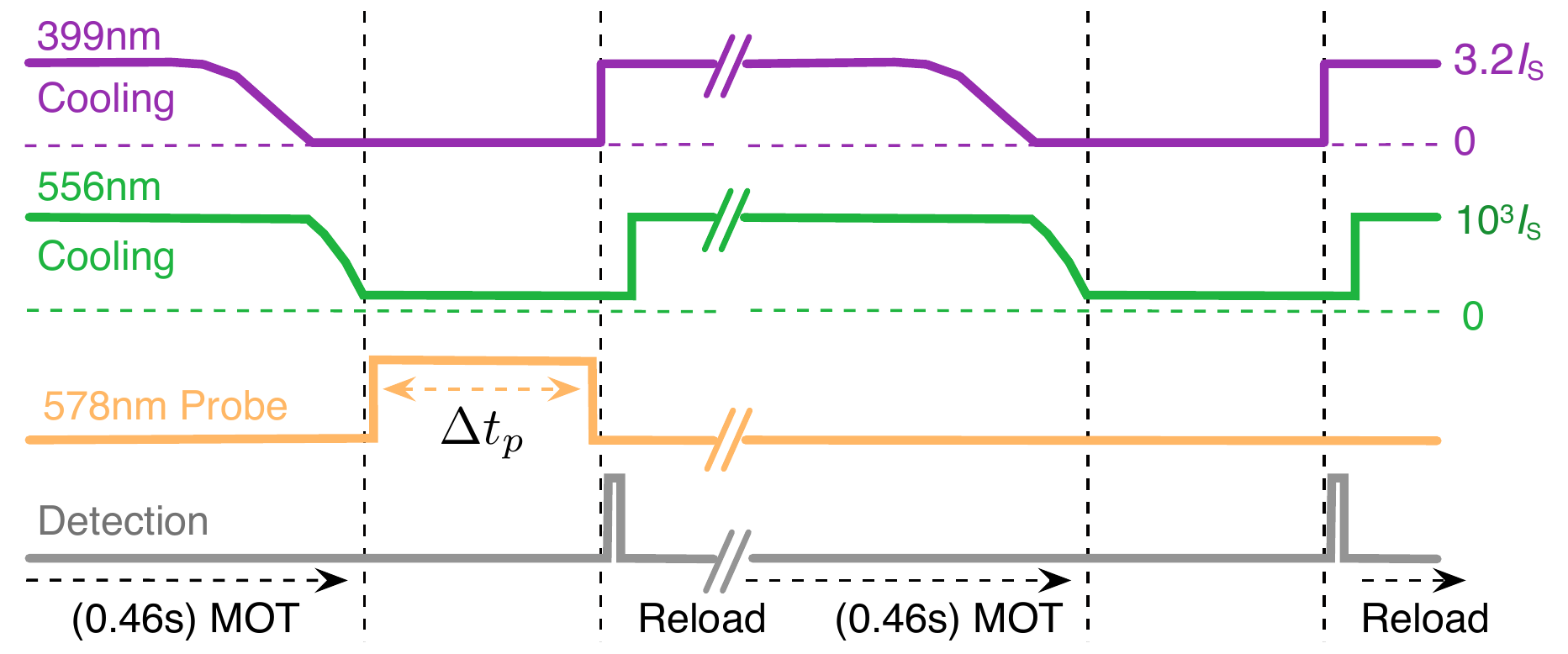}
        \caption{\small Event sequence for  the light fields.  
         The 578\,nm probe remains off  every second cycle. The duration $\Delta t_p=40$\,ms.  The saturation intensities are $I_S^{(399)}=595$\,W\,m$^{-2}$ and $I_S^{(556)}=1.4$\,W\,m$^{-2}$  for the \fastT\ and \coolingT\ transitions, respectively.  The frequency detuning of the cooling lasers is also varied to optimize signal level. 
        }  %
             \label{SequenceYb}
    \end{center}
\end{figure}

\begin{figure}[h!]
    \begin{center}
        \includegraphics[width=0.5\textwidth]{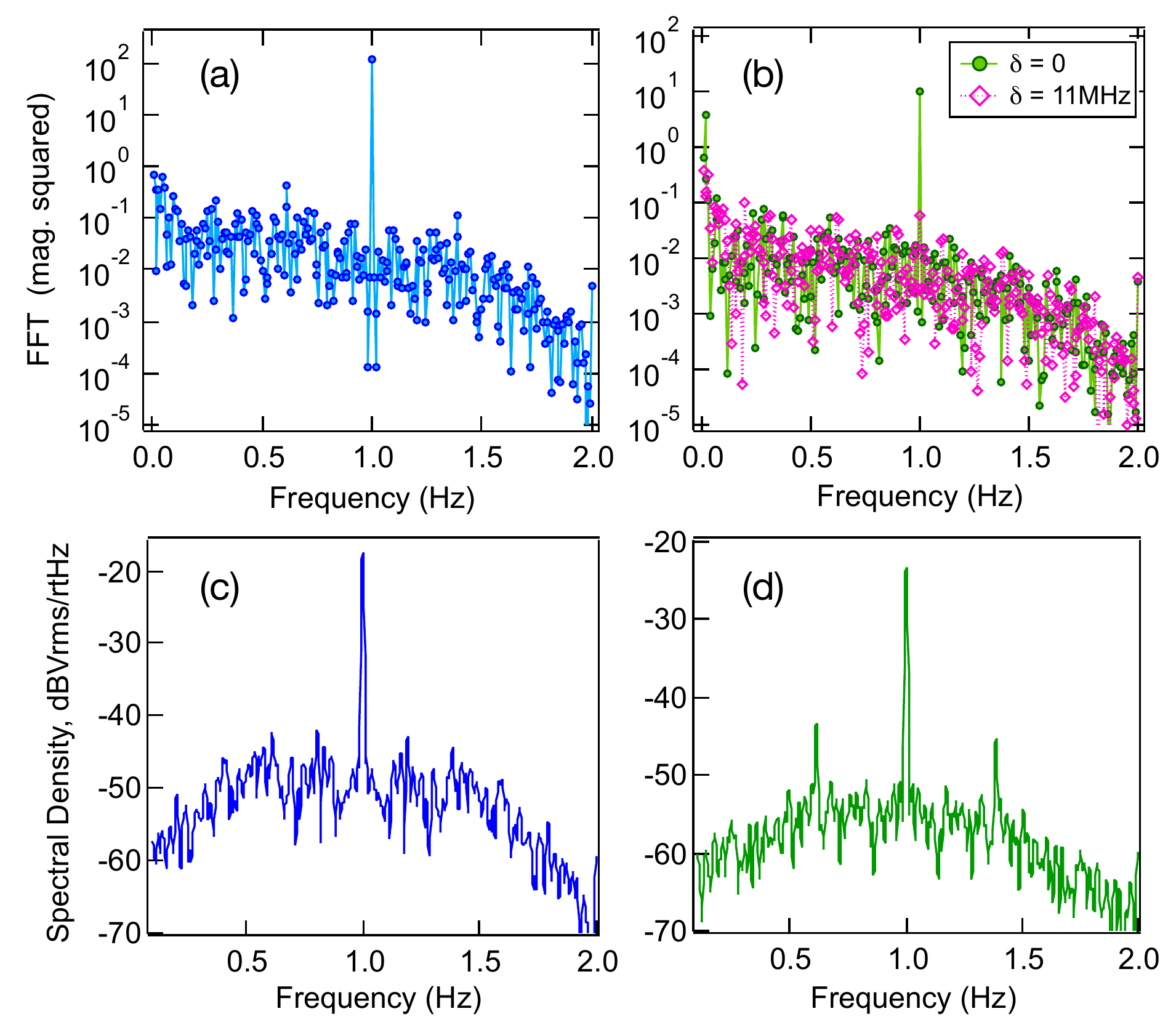}
        \caption{ \small Upper panel:  Fast Fourier transforms of the BCA output when the 578\,nm light is near the center of resonance for (a) \Yb, and (b) \Ybthree\ (circles).  Lower panel:  equivalent spectra recorded with an FFT spectrum analyser for (c) \Yb, and (d) \Ybthree. Plot (b) also shows the FFT when the 578nm light is offset from resonance by 11\,MHz (diamonds). 
        }   %
             \label{FFTs}
    \end{center}
\end{figure}

The event sequence is presented in Fig.~\ref{SequenceYb}.  Each cycle has a duration of 0.5\,s, the majority of which is used to load atoms into the MOT.   The 399\,nm light is ramped down over 20\,ms  during which time the second stage cooling with 556\,nm light takes effect. The 556\,nm intensity is also ramped down to reduce the temperature of the atoms and  minimize the ac Stark shift. The magnetic quadrupole field  is held fixed with a $z$-axis gradient of 0.42\,T$\cdot$m$^{-1}$.   The yellow 578\,nm light pulse has a period of 40\,ms to interact with the atoms, after which the  399\,nm is switched back on and detection of the ground state atoms is made (and MOT loading recommences).  Unless otherwise stated, the maximum intensity of the 578\,nm light was 5\,kW$\cdot$m$^{-2}$ with a corresponding Rabi frequency, $\Omega_R$, of \si4\,kHz (representing \si\ 1/50$^{th}$ of the transition linewidth).  
 A 1\,Hz modulation in the population transfer arises by  applying the 578\,nm light every second cycle. 
Fluorescence at 399\,nm is filtered both spatially and spectrally before detection with a photomultiplier tube (PMT, Hamamatsu H10492-001)~\cite{Kos2014}.  The PMT signal is received by a boxcar averager (BCA, SRS SR250), which outputs to an Agilent 34970A unit for data acquisition. 
The BCA is triggered by the main control sequence program so that detection is made within 300\,\us\ of the 399\,nm  light  resuming. 
A single measurement involves logging $N$ samples of the BCA output every 0.25\,s, such that the total acquisition time gives an integer number of modulation cycles. 
  A FFT of the time record (with Igor Pro software) reveals a signal at the modulation frequency if the probe laser frequency is on resonance.  %
  The upper panel of Fig.~\ref{FFTs} shows  examples of  FFTs (magnitude squared) for the (a) \Yb\  and (b) \Ybthree\ clock transitions, where $N=500$  in both cases (over 125\,s). These are maximum strength signals recorded very close to line center.  The SNR at the 1\,Hz modulation is approximately $6\times10^3$ for \Yb\ and  just under $2\times10^3$ for \Ybthree\ (we will denote this $SNR_\mathrm{fft\_or}$ $-$ `or' for on-resonance).  The number of atoms in the 556\,nm MOT  generating the signal was \si\  $5\times10^5$  and \si\ $2\times10^5$, respectively (the transfer fraction to $^1P_0$ was $\sim30$\%).  The Yb oven was  deliberately set at a relatively low temperature of 380\degC. 
 In an alternative means of obtaining the spectrum, the BCA output was sent to a FFT spectrum analyser (HP89410A) set with a resolution bandwidth  of 10\,mHz and a Hanning window.  The lower panel of Fig.~\ref{FFTs} shows the resultant spectra.
Figures~\ref{FFTs}(c) and  \ref{FFTs}(d) are for \Yb\ and \Ybthree, respectively. 
  In (d) the oven was at a higher temperature of 410\degC.  The SNRs at 1\,Hz are consistent with those in the upper panel.  The signals at 0.61\,Hz and 1.38\,Hz are unrelated to the 578\,nm light.   The 1.38\,Hz modulation appears to originate  from the Ti:sapphire laser and the 0.61\,Hz is the same modulation  shifted in frequency through aliasing. 
  
\begin{figure}[h!]
    \begin{center}
        \includegraphics[width=0.49\textwidth]{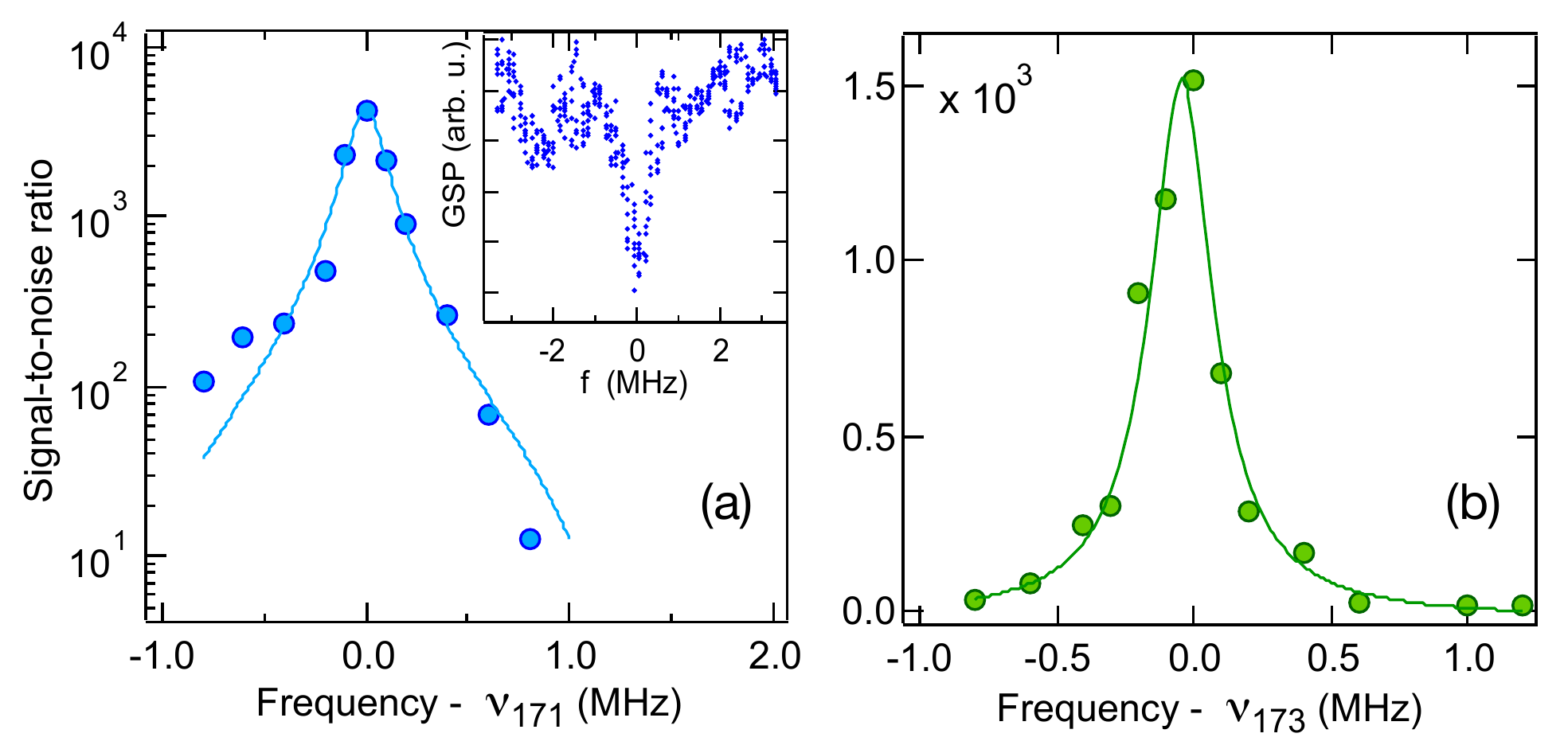}  %
        \caption{ \small SNR (magnitude squared) versus optical frequency  for the \clockT\ line in (a) \Yb\ and (b) \Ybthree\ by use of the FFT and cyclic modulation.  
          The inset shows a spectrum for \Yb\ with a conventional scan. GSP, ground state population. 
        }  
             \label{Spectra}  
    \end{center}
\end{figure}

A full line spectrum of the \clockT\ transition is produced by stepping the frequency of  AOM-1 (with a fixed frequency for each time record) and extracting the SNR from the FFTs.    
 Figures~\ref{Spectra}(a) and \ref{Spectra}(b) show  the \clockT\ line spectra for the \Yb\ and \Ybthree, respectively.   Each data point is determined from a separate FFT.  Figure~\ref{Spectra}(a) is shown with a logarithmic scale to illustrate the sensitivity in the wings of the profile.  The absolute frequency is determined from Eq.~\ref{EqComb}.     The frequencies subtracted from the line centers  are $\nu_{171}=518\,295\,836\,590.9$\,kHz~\cite{BIPM2017} and   $\nu_{173}=518\,294\,576\,847.6$\,kHz~\cite{Hoy2005}. The inset of Fig.~\ref{Spectra}(a) shows the spectrum by use of a conventional frequency sweep for \Yb\ with a similar number of atoms. In this case  the 578\,nm pulse is applied every cycle of the event sequence, unlike in Fig.~\ref{SequenceYb}.  Here the total sampling time was 140\,s 
 with the BCA set to 3-sample averaging.    Note, the noise here is not white frequency noise.  
 
 To compare the detection sensitivity of the two methods we introduce a sensitivity index $d' = (\mu_s-\mu_n)/\sigma_n$, where $\mu_s$ is the mean of the signal level, $\mu_n$ is the mean of the noise level, and $\sigma_n$ is the standard deviation of the noise~\cite{Wic2002}.    Here $\mu_s$ corresponds to the on-resonance signal.  
 For the FFT approach the units are magnitude-squared, therefore,  $d'_\mathrm{fft}= \sqrt(SNR'_\mathrm{fft\_or})$.
In the case of the conventional sweep we obtain $d'_\mathrm{swp}=3.3$, hence $d'_\mathrm{fft\_or}/d'_\mathrm{swp}\approx 23$; or in power units, the gain in sensitivity is \si\ 27\,dB.   Hence the means of  detecting the transition over similar time scales is very much enhanced through the FFT approach.

The spectra in Figs.~\ref{Spectra}(a) and Fig.~\ref{Spectra}(b) have their own SNR. 
  In this case we do not use the noise floor of the FFT to set the noise level.   More appropriate is the rms of the residuals to the line shape fits.  Here the SNR is 38  and  16  for \Yb\ and \Ybthree, respectively.       The conventional scan for  \Ybthree\ produced no evident  transition over the 140\,s of sampling, due to the lower number of atoms trapped. 
 The total sample time for Figs.~\ref{Spectra}(a) and Fig.~\ref{Spectra}(b) was 1500\,s and 1300\,s, respectively, so while there is a significant gain in SNR, the total sampling time increased by \si\ 10 fold.
%
From the Lorentzian line shape fits (to suit the wings) the FWHM for traces (a) and (b) are 210\,kHz and 290\,kHz, respectively.  The width for \Yb\ is consistent with the temperature of the atoms determined by imaging of ballistic expansion, $T\sim50$\,\uK\ \cite{Kos2014}.  For \Ybthree\ there may be a linewidth contribution from  laser noise~\cite{Rey2019}, since the  FVC lock to the comb was used rather than the ultrastable cavity. 

\begin{figure}[h!]
    \begin{center}
        \includegraphics[width=0.49\textwidth]{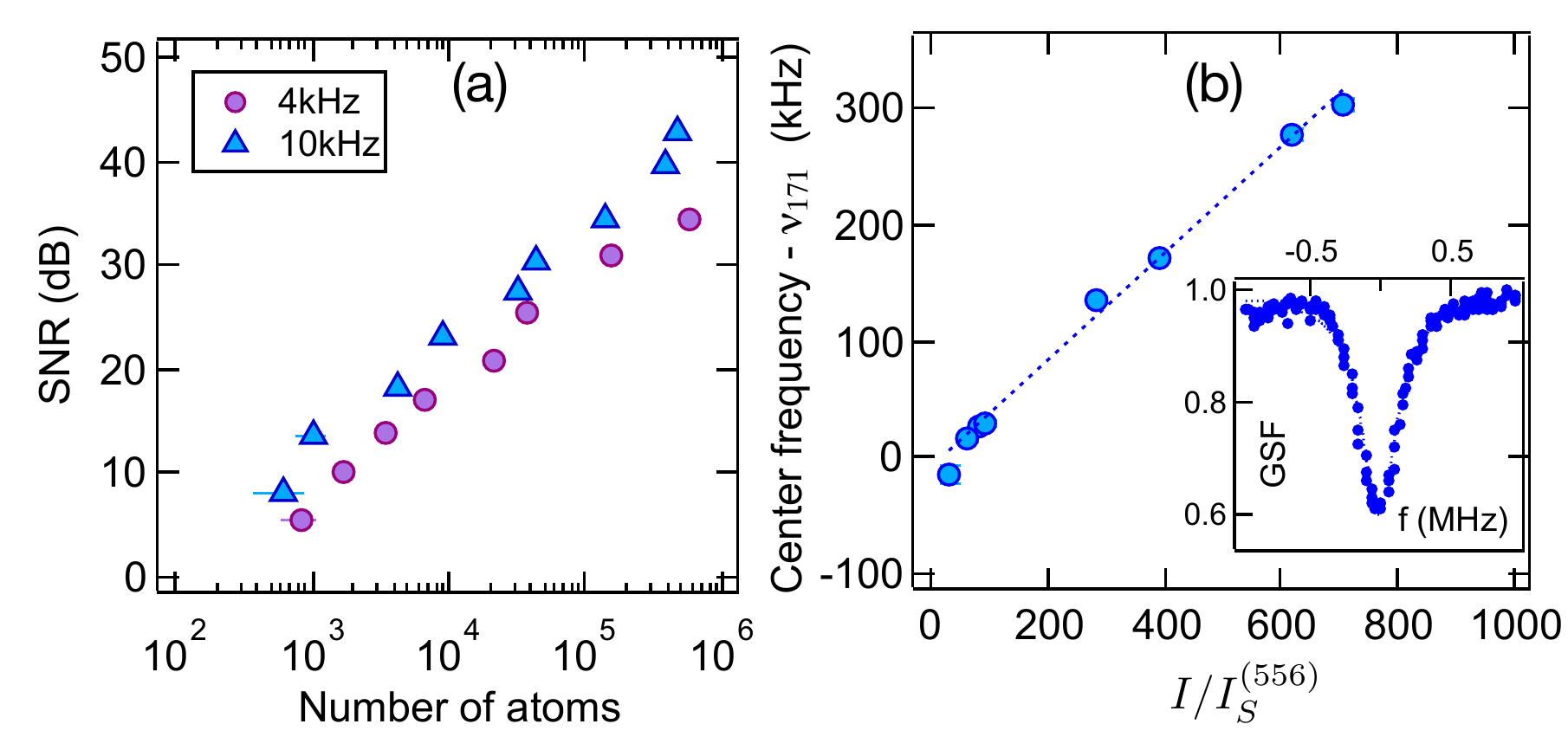}  %
        \caption{ \small (a)  SNR of the 1\,Hz FFT signal, on resonance, versus atom number for two values of $\Omega_R$. (b) Frequency of the   \Yb\ \clockT\  transition versus 556\,nm light intensity. 
     Inset: spectrum  of the   \Yb\ clock line  after optimization. GSF, ground state fraction.   
        }  
             \label{Spectrum171}
    \end{center}
\end{figure}

Figure~\ref{Spectrum171}(a) shows the on-resonance 1\,Hz SNR  
 versus the number of atoms contributing to the signal for \Yb\ with $\Omega_R=4$\,kHz and $10$\,kHz (varied through 578\,nm intensity). The number of atoms is determined by use of $N_{a} = V/( h \nu G f \gamma_{p})$, where $V$ is the dc voltage from the PMT (background subtracted), $h$ is Planck's constant, $\nu$ is the frequency of  the 399\,nm light, $G$ is the gain of the PMT (V/W), $f$ is the collected fraction of fluorescence, and $\gamma_{p}$ is the photon scattering rate.
The scattering rate follows from: 
   $\gamma_{p} = s_{0}\gamma/ (1+s_{0}+(2 \delta/\gamma)^{2})/2$, where $s_{0}$ is the 399\,nm intensity normalized by the saturation intensity, $I_S^{(399)}$, $\delta$ is the frequency detuning and $\gamma$ is the natural linewidth of the \fastT\ transition~\cite{Met1999}.  
 The atom number was varied by changing the current through the Zeeman slower coils.  With $\sim10^3$ atoms a  Fourier signal $\gtrsim10$\,dB SNR remains apparent when $\Omega_R=10$\,kHz.  The sensitivity increases with  578\,nm intensity.  Note, our  fluorescence collection efficiency is only 0.55\,\%.

 Once an atomic line is found, the modulated BCA output can aid  optimization, such as improving the probe beam overlap with the atom cloud.   An example of the \Yb\ clock line recorded with the conventional scan is shown in the inset of Fig.~\ref{Spectrum171}(b) after such an optimization (and $T_\mathrm{oven}=410$\degC).  The green light intensity during the 578\,nm pulse was  $I/I_S^{(556)}=140$.    By repeating with different levels of 556\,nm light the ac-Stark shift becomes evident, as seen in Fig.~\ref{Spectrum171}(b).   The line center for zero light intensity is within 9\,kHz of  previous reports of the transition frequency~\cite{Lud2015,BIPM2017}. The gradient of the line fit for the ac-Stark shift is 0.19(3)\,kHz$\cdot$W$^{-1}\cdot$m$^2$, where the uncertainty is dominated by that of the light intensity. 

In summary, we have demonstrated a sensitive method of detecting weak optical transitions in cold atoms  that rely on cyclic routines.  In terms of detecting the presence of a transition, the sensitivity index is increased 20-fold 
compared to conventional line scanning for similar measurement times. 
One can detect (and resolve) a clock transition in an optical lattice with  $\sim10^3$ atoms~\cite{Yi2011}, but to our knowledge  the detection of such a transition with $10^3$ atoms in a MOT  has not been previously demonstrated. With regard to  line spectra, the increased SNR comes at the expense of a 10-fold increase in sampling time, which places a restriction on   $\Delta f_S/\Delta \nu$ (mentioned in the opening).  
An upper limit for $\Delta f_S/\Delta \nu$  may be \si\ 20 for practical purposes, but it depends on factors such as cycle time and how well search procedures can be automated.   

\vspace{-0.1cm}

\begin{acknowledgments}
This work was supported by the \ARC's grant CE170100009. 
J.~S. acknowledges support from the University of Western Australia's Winthrop Scholarship, and St Catherine's College.  
We  thank F.~van~Kann and L.~Nenadovi\'c for proofreading the manuscript.\\
\end{acknowledgments}

\section*{Data Availability}
The data that support the findings of this study are available from the corresponding author upon  request.

\nocite{*}

\begin{thebibliography}{10}
\bibitem{Comment1} 
  Lower temperature may be obtained with optimal MOT beam alignment.
\end{thebibliography}

\section*{References}  

\setstretch{1.8}

\end{document}